\documentclass[structabstract]{aa}  
\usepackage{graphicx}
\usepackage{longtable}
\usepackage{natbib}
\usepackage{txfonts}
\def\gsim{\mathrel{\hbox{\rlap{\hbox{\lower4pt\hbox{$\sim$}}}\hbox{$>$}}}}
\def\lsim{\mathrel{\hbox{\rlap{\hbox{\lower4pt\hbox{$\sim$}}}\hbox{$<$}}}}
\def\Msun{\hbox{$\rm\thinspace M_{\odot}$}}

\begin{document}
   \title{X-ray variability of 104 active galactic nuclei}
   \subtitle{\emph{XMM-Newton} power-spectrum density profiles}
   \author{O. Gonz\'alez-Mart\'in\inst{1,2,3}\thanks{Juan de la Cierva fellow (\email{omairagm@iac.es})} \and S. Vaughan\inst{4}
	  }

   \institute{ Instituto de Astrof\'isica de Canarias (IAC), C/V\'ia L\'actea, s/n, E-38205 La Laguna, Spain \and Departamento de Astrof\'isica, Universidad de La Laguna (ULL), E-38205 La Laguna, Spain \and IESL, Foundation for Research and Technology, 711 10, Heraklion, Crete, Greece \and X-ray Astronomy Group, Department of Physics and Astronomy, Leicester University, Leicester LE1 7RH, UK
             }

   \date{Received ?? 2012; accepted ??, 2012}
% \abstract{}{}{}{}{} 
% 5 {} token are mandatory
 
\abstract  
   {Active galactic nuclei (AGN), powered by accretion onto supermassive black holes (SMBHs), are thought to be scaled up versions of Galactic black hole X-ray binaries (BH-XRBs). In the past few years evidence of such correspondence include similarities in the broadband shape of the X-ray variability power spectra, with characteristic bend times-scales scaling with mass. }
   {The aim of this project is to characterize the X-ray temporal properties of a sample of AGN to study the connection among different classes of AGN and their connection with BH-XRBs.}
   {We have performed a uniform analysis of the power spectrum densities (PSDs) of $104$ nearby (z$\rm{<0.4}$) AGN using $209$ \emph{XMM-Newton}/pn observations. These PSDs span $\rm{\simeq 3}$ decades in temporal frequencies, ranging from minutes to days.  The PSDs have been estimated in three energy bands: $0.2-10$ (total), $0.2-2$ (soft), and $2-10$ keV (hard). The sample comprises $61$ Type-1 AGN, $21$ Type-2 AGN, $15$ NLSy1, and $7$ BLLACS. We have fitted each PSD to two models: (1) a single power-law model and (2) a bending power-law model.}
   {Among the entire sample, 72\% show significant variability in at least one of the three bands tested. A high percentage of low-luminosity AGN do not show any significant variability (86\% of LINERs). The PSD of the majority of the variable AGN was well described by a simple power-law with a mean index of $\rm{\alpha =}$      2.01$\rm{\pm}$      0.01
. In $15$ sources we found that the bending power law model was preferred with a mean slope of $\rm{\alpha =}$      3.08$\rm{\pm}$      0.04
 and a mean bend frequency of $\langle \nu_{b} \rangle \simeq 2\times 10^{-4}$ Hz. Only KUG\,1031+398 (RE\,J1034+396) shows evidence for quasi-periodic oscillations. The `fundamental plane' relating variability timescale, black hole mass, and luminosity is demonstrated using the new X-ray timing results presented here together with a compilation of the previously detected timescales from the literature.
    }
   {Both quantitative (i.e. scaling with BH mass) and qualitative (overall PSD shapes) found in this sample of AGN are in agreement with the expectations for the SMBHs and BH-XRBs being the same phenomenon scaled-up with the size of the BH. The steep PSD slopes above the high frequency bend bear a closer resemblance to those of the `soft/thermal dominated' BH-XRB states than other states.
   }
 

\maketitle

% =============================================================================
% =============================================================================

\section{Introduction}\label{sec:intro}

The dynamics of accretion around BHs should be similar for different masses, with characteristic size scales and hence timescales simply scaling with their mass (at a given accretion rate relative to the Eddington limit). Therefore, we expect to find similarities between the properties of luminous accretion flows around stellar mass BHs in X-ray binaries (BH-XRBs; $M_{BH} \sim 10$ \Msun) and supermassive black holes (SMBHs) in Active Galactic Nuclei (AGN; with $M_{BH} \sim 10^6 - 10^9$ \Msun). However, the details and reliability of the proposed scaling relations are currently not well understood \citep[see][for a review]{Merloni03,Koerding07,McHardy10}.

X-ray variability -- thought to originate in the innermost regions of the accretion flow -- is an important aspect of the AGN-XRB connection \citep[e.g.][]{Uttley02,Markowitz03,Vaughan03B,McHardy06,McHardy10, Vaughan11}. Both AGN and BH-XRBs show `red noise' power spectra (or power spectral density, PSD) that decrease steeply at high frequencies (short timescales) as a power law, $P(\nu) \sim \nu^{-\alpha}$ (where $\nu$ is temporal frequency), typically with $\alpha \approx 2$. Below some characteristic frequency $\nu_{b}$ the PSDs flatten, and these bend frequencies scale approximately inversely with the BH mass from BH-XRBs to AGN. However, the PSDs of BH-XRBs depend on the `state' in which the system is observed.  In the \emph{soft state} the PSD is usually well described by a simple bending power-law, with a slope $\alpha > 2$ above some bend frequency $\nu_{\rm b}$, and a slope of $\alpha \approx 1$ extending unbroken to much lower frequencies \citep{Cui97}. In the \emph{hard state} the PSDs are generally more complex, with additional bends and peaks at lower frequencies, and are usually modeled as a mixture of zero-centred Lorentzian components \citep[see e.g.][]{Remillard06, Wilms06}. In addition there are transitional states that display a range of timing behavior. The accretion `states' of AGN are not clearly understood \citep{Koerding06,Fender06} but there is some evidence to support the idea that luminous, radio-quiet AGN represent supermassive analogues of BH-XRB in the `soft states' \citep{Uttley02,McHardy04,Uttley05,Vaughan11}.

One striking and common property of the X-ray variability of BH-XRBs is the prominent quasi-periodic oscillations \citep[QPOs, see e.g.][]{Remillard06}, which are seen as strong, narrow peaks in the PSDs. Until recently there were no robust claims of QPOs in AGN \citep[see e.g.][]{Vaughan05B}. This changed when \citet{Gierlinski08} reported a QPO for the Narrow-line Seyfert 1 (NLSy1) RE\,J1034+396, which remains the only case to date for an AGN \citep[see also][]{Middleton09, Vaughan10}.

The main purpose for the present paper is to characterize the PSD of a large sample of AGN including a range of AGN subclasses. With this large sample we address some outstanding problems such as: what is the most common PSD shape for AGN? how does this compare to BH-XRBs? How does the PSD vary with AGN properties (BH mass, AGN subtype or luminosity)? Are there any other strong candidate of QPO among the AGN? We have made use of the extensive \emph{XMM-Newton} archive to obtain a sample of $104$ AGN with $209$ observations of at least $40$ ksec. This is the largest sample of PSDs of AGN ever analyzed.

% =============================================================================
% =============================================================================

\section{The sample and the data}\label{sec:sample}

We have selected observations of AGN from the \emph{XMM-Newton} public archives until February 2012. Note that this should include all the sources within the 2XMMi DR3 catalogue \citep{Watson09}. The AGN are all identified in the catalogue of Quasars and AGN by \citet{Veron10}\footnote{http://heasarc.nasa.gov/W3Browse/all/veroncat. html} with $z < 0.4$. We have selected data according to the following inclusion criteria: (1) include one or more members of the AGN sample (search radius $9$ arcmin from the centre of the FOV), and (2) observation duration $T > 40$ ksec. The second criteria ensures a reasonable frequency range for the PSD estimation (the lowest observable frequency scales is $1/T$). The final sample comprises $209$ observations and $104$ distinct AGN.

Table \ref{tab:properties} gives details of the sample members and relevant \emph{XMM-Newton} observations. The optical classification (Col. 3) is taken from \citet{Veron10}. Three objects (NGC\,4636, NGC\,4736, and NGC\,6251) were re-classified into the LINER class according to \citet{Gonzalez-Martin09}. The sample comprises $61$ Type-1 AGN ($3$ QSOs, $54$ Seyferts, and $4$ LINERs), $21$ Type-2 AGN ($11$ Seyferts and $10$ LINERs), $15$ NLSy1, and $7$ BLLACs. Note that we consider S1i class in the catalogue (broad Paschen lines observed in the infrared) as Type-1 Seyfert and S1h (broad polarized Balmer lines detected) as Type-2 Seyferts. The redshifts given in Col.~4 (Tab. \ref{tab:properties}) were taken from the NASA/IPAC extragalactic database (NED\footnote{http://nedwww.ipac.caltech.edu}) using the redshift-independent distance when it was available. The cosmology assumed was $H_{0} = 75$ km s$^{-1}$ Mpc$^{-1}$. BH masses ($M_{BH}$) and bolometric luminosity ($L_{bol}$) reported in the literature are shown in Cols.~5 and 6 together with the corresponding reference (Col. 7). We included the reverberation mapping estimate if available and the velocity dispersion estimate using the correlation reported by \citet{Tremaine02} otherwise. The velocity dispersions are taken from HyperLeda database\footnote{http://leda.univ-lyon1.fr}.

% =============================================================================
% =============================================================================

\section{Data reduction}\label{sec:datared}

\emph{XMM-Newton} data were processed using the {\sc SAS} (v10.0.2) and the most recent calibration files available. In this paper, only the EPIC pn data \citep{Struder01} were analyzed; the pn data have a higher count rate and lower pile-up distortion than the MOS data \citep[the effect of pile-up on PSD estimation is discussed by][]{Tomsick04}.

Time intervals of quiescent particle background were determined through an algorithm that maximizes the signal-to-noise ratio of the net source spectrum by applying different constant count rate threshold on time series of single-event, E$\rm{>}$10 keV events, from the full field-of-view \citep{Piconcelli04}.

The AGN sky coordinates were retrieved from NED\footnote{The NASA/IPAC Extragalactic Database (NED): {\tt http://ned.ipac.caltech.edu/}} and source counts in each case were accumulated from a circular aperture of radius $40$ arcsec\footnote{This includes approximately $90$ per cent of the PSF for an on-axis source with the EPIC pn instrument.}). The background events were selected from a source-free circular region on the same CCD chip as the source. We selected only single and double pixel events (i.e. {\tt PATTERN$==$0$-$4}). Bad pixels and events too close to the edges of the CCD chip were rejected (using the standard {\tt FLAG}$==$0 inclusion criterion).

Source and background light curves were extracted (using {\sc evselect}) in $100$s time bins, in three energy ranges: $0.5-10$ keV (total band, T), $0.5-2$ keV (soft band, S), and $2-10$ keV (hard band, H). These were screened for high background and flaring activity \citep[see][for a detailed explanation of the procedure]{Gonzalez-Martin11}. High background periods at the beginning and/or end of an observation were removed. Where short periods of high background occurred in the middle of observations we excised the bad data and used the expected values derived by a linear interpolation of the neighbor data points to replace the missing data. This replacement is only done when less than 5 consecutive points need to be interpolated. For longer periods, the lightcurve is truncated selecting the largest consecutive period. Note that drop outs are less than 1\% of the duration of the light-curve in all the observations presented here. Only light curves with remaining good durations longer than $40$ ks were considered for further analysis. The net exposure times are listed in Col. 8 of table \ref{tab:properties}. We also extracted the EPIC-pn spectra, with response matrix and auxiliary files generated using {\sc rmfgen} and {\sc arfgen}, respectively. 

% =============================================================================
% =============================================================================

\section{Data analysis}\label{sec:analysis}

	\subsection{Hard band 2-10 keV luminosity estimates}\label{sec:lum}

The $2-10$ keV luminosity was estimate by fitting the $2-10$ keV spectrum using an absorbed power-law model (using {\sc xspec} version 12.6.9q). For the Galactic absorption we used the hydrogen column density ($N_H$) estimates from the H{\sc I} maps of \citet{Dickey90}. Additional absorption was also considered to account for the curvature of the spectrum of some AGN. The free parameters in the model were the intrinsic $N_H$, spectral index, and the power-law normalization. We did not attempt a more detailed spectral analysis, as we required only reasonable estimates of the X-ray luminosity. This method was already used by \citet{Ho01} to estimate the luminosity in sources with low signal-to-noise data. The resulting $2-10$ keV luminosities are given in Col. 10 of Table \ref{tab:properties}.

We have compared these luminosities with those for the same objects reported by \citet[][and references therein] {Gonzalez-Martin09, Panessa06, O'Neill05}. LLAGN (i.e. Type-2 Seyferts and LINERs) show a very good agreement while Type-1 Seyferts, QSOs, and NLSy1 show the highest discrepancies, as expected due to the high long-term variability. NLSy1 show strong soft excesses interpreted as optically thick material seen in reflection \citep{Crummy06} or optically thin material seen in absorption \citep{Gierlinski04}. This soft-excess shows long-term variability \citep[see][and references therein]{Boller96}.

\begin{figure*}
\begin{center}
\includegraphics[width=0.7\columnwidth,angle=90]{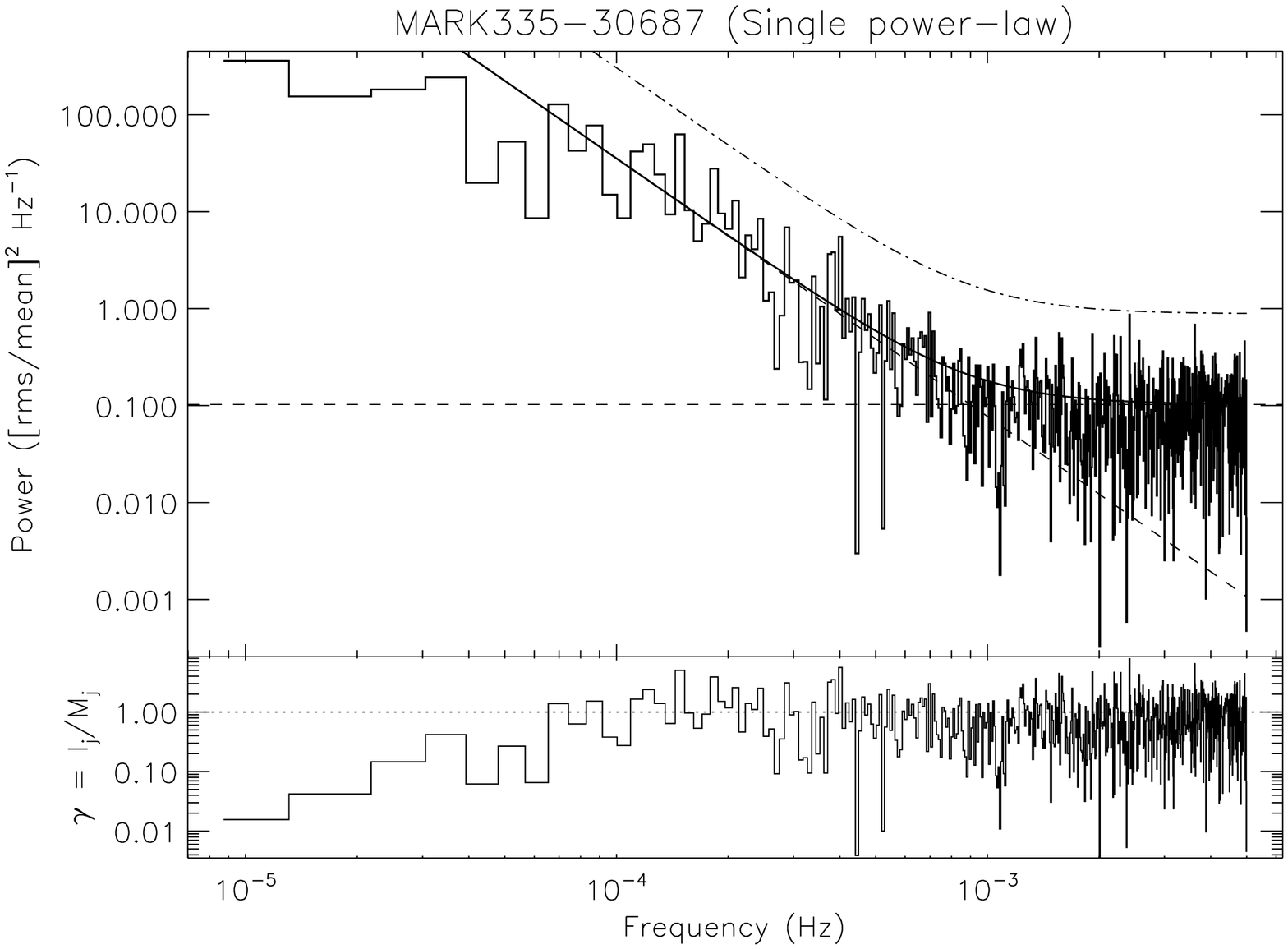}
\includegraphics[width=0.7\columnwidth,angle=90]{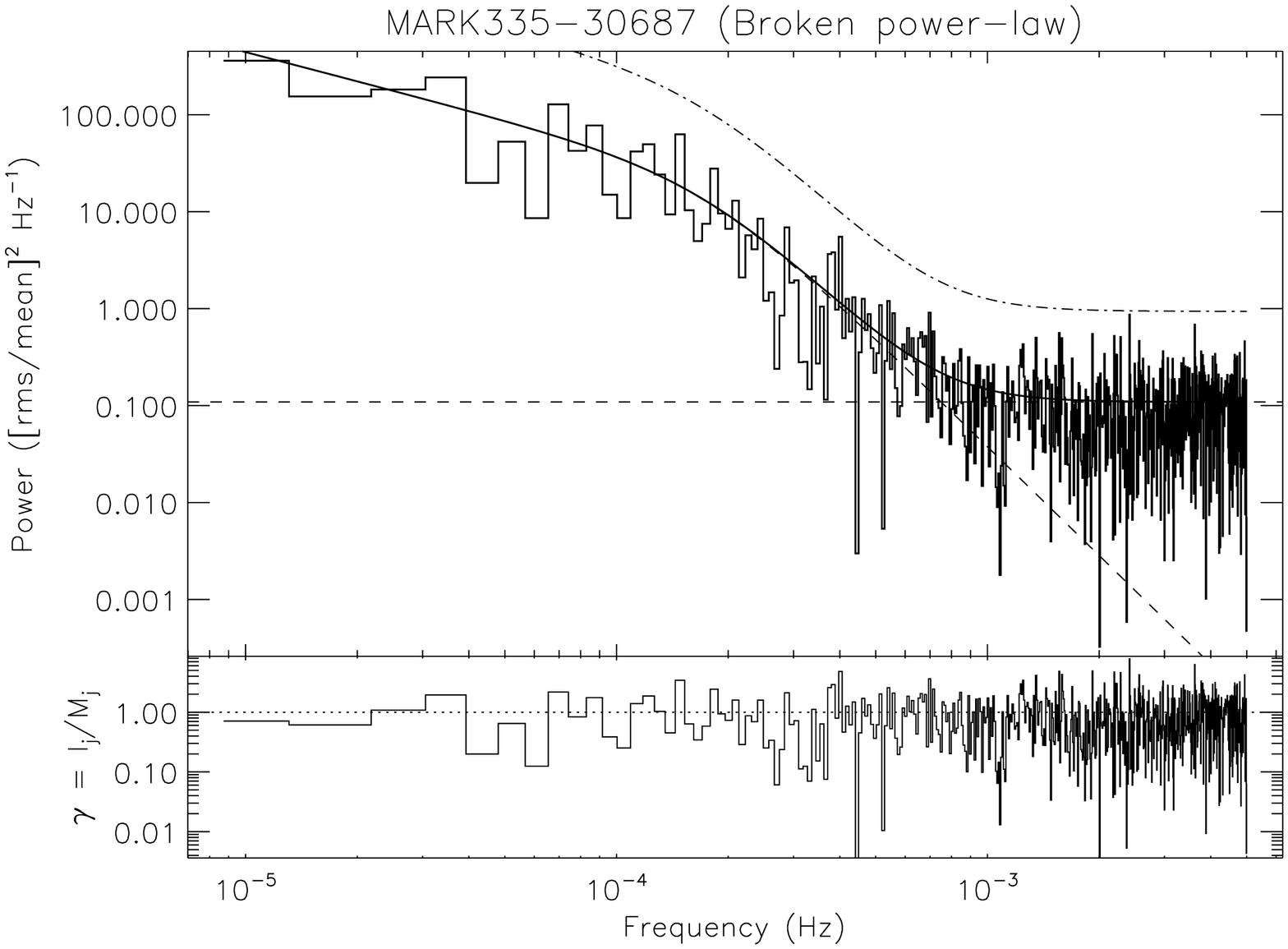}
\setcounter{figure}{0}
\caption{PSDs fits (continuous line) to Model A (left) and Model B (right) for the Mrk\,335 data (ObsID 306870101) using the broad ($0.2-10$ keV) energy band. The  dashed lines shows the two components of the model: constant Poisson noise and the source PSD model (power-law, left; bending power-law, right). The dot-dashed line shows the `global' 90\% confidence limit use to flag QPO candidates. Appendix B shows the corresponding figures for the complete sample.}
\label{fig:PSDs}
\end{center}
\end{figure*}

	\subsection{Power spectrum estimation}\label{sec:fit}

The PSD gives the distribution of variability power (amplitude squared) as a function of the temporal frequency. The standard method for estimating the PSD is by calculating the periodogram \citep{Priestley81,Percival93,Vaughan03A}. Fig. \ref{fig:PSDs} shows the periodogram  for Mrk\,335. All the periodograms of the sample are shown in the electronic edition of the paper. We use the ``(rms/mean)$^2$'' normalisation throughout \citep[see][and references therein]{Vaughan03A}.

The periodogram data were fitted using the maximum likelihood method discussed in \citet[][hereinafter V10]{Vaughan10} and \cite{Barret12}. For a given parametric model $P(\nu; \theta)$, the best-fitting model parameters, $\theta$, were found by maximizing the likelihood function, which in practice was done by minimizing the following fit statistic
\begin{equation} 
S     = 2 \sum_{j=1}^{N/2} \left\{ \frac{I_j}{P_j} + \log P_j \right\} 
\end{equation} 
This is (twice) the minus log-likelihood, where $I_j$ and $P_j$ are the observed periodogram and the model spectral density at Fourier frequency $\nu_j$, respectively. Confidence intervals on each model parameter were estimated by finding the set of parameter values for which $\Delta S = S(\theta) - S_{\rm min} \le 1.0$, this corresponds to $68.3$\% intervals \citep[this closely parallels the $\Delta C$ method discussed by][]{Cash79}. See also \cite{Vaughan05}.

Following the previous work on the PSDs of AGN we used two models: \begin{itemize} \item \underline{Model A}: simple power law plus constant (denoted $H_{0}$ by V10). 
\begin{equation} \label{eqn:m0}
P(\nu) = N \nu^{-\alpha} + C 
\end{equation} 
The model has three parameters: $N$, the power law normalization; $\alpha$, the spectral index; $C$, an additive constant to account for Poisson noise.

  \item \underline{Model B}: a bending power law plus constant (denoted $H_{1}$ by V10). 
  \begin{equation} \label{eqn:m1} 
  P(\nu) = N \nu^{-1} \left( 1 + \left\{ \frac{\nu}{\nu_{b}} \right\}^{\alpha-1} \right)^{-1} + C 
  \end{equation} 
  The free parameters for this model are: the normalization $N$, the spectral index above the bend $\alpha$, the bend frequency $\nu_{b}$, and the constant $C$. 
    \end{itemize} 

These \emph{XMM-Newton} data are relatively insensitive to the exact low frequency index, and we assume the typical value of $1$ found from long-term X-ray monitoring studies \citep[e.g.][]{Uttley02, Markowitz03, McHardy04, McHardy06}. The lowest frequencies available with these \emph{XMM-Newton} data are $\sim 0.7 - 2.5 \times 10^{-5}$ Hz (set by the minimum and maximum duration of continuous observations, $40$ and $\sim 130$ ks respectively). The typical high frequency bend for a low mass Seyfert galaxy occurs at around $\nu_{b} \sim few 10^{-4}$ Hz, which implies only $\sim 10-40$ Fourier frequencies below $\nu_{b}$, usually insufficient to obtain a precisely constrained power law index.

Temporally close observations of the same objects were analyzed together to better constrain the final parameters (the periodograms were fitted simultaneously with all parameters except $C$ tried between the datasets). Note that we consider as temporally close observations those coming from the same observing run according to the first 6 digits of the ObsID. The resulting parameters are shown in tables \ref{tab:psd0} and \ref{tab:psd1}.

The Likelihood Ratio Test (LRT) was used to select between the two models. The simpler model (A) is preferred in the absence of a strong preference for the more complex model (B). For reasons explained by \citet{Protassov02} and \citet{Freeman99} the LRT is not well calibrated when when the null values of the additional parameter of the more complex model are not well defined, as is the case here (e.g. the null value for the bend frequency in Model B is not well defined in this sense). An  accurate calibration of this test using posterior predictive checks can be computationally expensive (see V10), so in this case we used the standard LRT with a $p < 0.01$ significance threshold as a crude but simple check for the presence of a bend in the PSD. The simulation results of V10 suggest that this criterion is conservative in the sence that the number of false positives is fewer than expected. Where a bend is detected (i.e. Model B selected) we parametrized the PSD amplitude at the bend frequency as $A_{psd} = \nu_{b} \times P(\nu_{b}) = N/2$ \citep[following][]{Papadakis04}, where $N$ is the normalisation (equation \ref{eqn:m1}), i.e. the peak in $\nu \times P(\nu)$ space.

The Poisson noise level, $C$, was usually considered a free parameter. The expected Poisson noise level was calculated assuming the standard formula \citep[e.g.][]{Uttley02, Vaughan03A} and this value was used in cases where the parameter $C$ was poorly constrained in the periodogram fitting (these observations are indicated by an asterisk in tables \ref{tab:psd0} and \ref{tab:psd1}).

As a crude but simple check for the presence of highly coherent oscillations (strictly or quasi-periodic) we identify the largest data/model outlier from each periodogram. This was then compared to the $\chi_2^2$ distribution expected for periodogram data to give a $p$-value. Candidate QPOs were flagged when a $p < 0.01$ criterion was met (after correcting for the number of frequencies in the periodogram). A similar test was applied to the periodogram data after smoothing using a three-point top-hat filter, i.e. testing the running mean of each triplet of adjacent points. This test improves sensitivity to QPOs that are broader than the frequency resolution of the data. These $p$-values are not correctly calibrated, for reasons explained in \citet{Vaughan10}, but such tests proved to be extremely efficient as a crude indicator of potentially interesting data.

There are two main types of bias that may effect Fourier-based power spectral analysis, 'aliasing' and 'leakage' \citep[see][]{Priestley81, Percival93, Deeter82, Uttley02}. Aliasing is negligible for the present data, which are contiguously sampled. Leakage may however cause intrinsically steep power spectra (power law indices $\alpha \gsim 2$) to show slopes close to $\alpha = 2$. This point is discussed further in section \ref{sec:leakage}.

% =============================================================================
% =============================================================================

\section{Results}\label{sec:results}

	 \subsection{Variability of the sample}

We considered an observation variable if there was an excess above the constant value expected from the Poisson noise in the periodogram. We report the detection of variability in Col.~9 of Table \ref{tab:properties} as: (T) for the total $0.5-10$ keV, (S) for the soft $0.5-2$ keV, and (H) for the hard $2-10$ keV energy bands; (N) indicates no variability detected in any band.

Seventy five out of the $104$ AGN (72\%) showed variability in at least in one of the three bands tested. Among them, $56$ showed significant variability in the total band, while three varied only in the hard band (NGC\,985, 2E\,0414+0057, and NGC\,4258). Twenty nine of the $104$ AGN (28\%) did not show significant variability in any of the three bands tested. These sources are: $12$ of the $14$ LINERs; $2$ of the $11$ Type-2 Seyferts; $12$ of $54$ Type-1 Seyferts; $2$ of the $3$ QSOs; and $1$ of the $7$ BL Lacs. All the NLSy1s and $78$\% of the Type-1 Seyferts were variable, but most of LINERs are not variable. Finally, five sources were variable only in some of the observations (namely, OJ\,+287, NGC\,2992, PG\,1116+215, Mrk\,279, Mrk\,205, 3C\,273, X\,Com, and RBS\,1399). The lack of detected variations in these cases can be attributed to the limited quality of the data.

	 \subsection{Model selection}\label{sec:psdfit}

Among the variable sources, we have been unable to constrain the parameters of Model A in six sources, and an additional nine have unconstrained values in a subset of the observations and/or energy bands (marked as $\rm{\clubsuit}$ in Table \ref{tab:psd0}). This is due  to the low number of bins in the PSD above the Poisson noise (e.g. see Fig. \ref{fig:PSDs} of NGC\,4151, ObdID = 112830201,  for the total energy band in the electronic edition). 

Model B is selected over Model A (using the LRT) in $17$ AGN ($44$ observations, see Tab. \ref{tab:psd1}). However, MS\,22549-3712 and Mrk\,586 gave very poorly constrained parameters for Model B. We consider these two sources as tentative detections of a bending PSD. Moreover, we have found a break in PKS\,0558-504 far from that reported by \citet{Papadakis10}. However, our bend is more likely a bump on the PSD or a broad QPO as claimed by \citet{Papadakis10}. All the objects for which Model B was selected are Type-1 AGN but the Circinus galaxy.

	\subsection{QPO detection}

The test for highly periodic oscillations found no candidates besides KUG\,1031+398 (better known as RE\,J1034+396). A QPO of a $\rm{\nu_{b}=2.6\times 10^{-4}}$ Hz was previously detected in this source by \citet{Gierlinski08}. The apparent significance of the QPO is lower than that reported by \citet{Gierlinski08}, for reasons discussed by V10.

A QPO candidate was previously reported for 3C\,273 by \citet{Espaillat08} using a wavelet technique to analyze $19$ observations of $10$ AGN obtained with \emph{XMM-Newton}. For 3C\,273 they used four \emph{XMM-Newton} observations, all of them included in our analysis. We found no indication of a QPO from these data. Another case of QPO reported in the literature is the blazar PKS\,2155-304. \citet{Lachowicz09} reported  a $4.6$ h QPO in the $0.3-10$ keV data (from ObsID 158961401). We have analyzed $8$ observations of this source, including this one; none showed indications of a QPO. \citet{Gaur10} analyzed all the \emph{XMM-Newton} archival data of this source and reported a possible QPO in only one observation. This observation is included in our analysis and the hint of the QPO around $\rm{1.7\times 10^{-4}}$ Hz can be seen in our PSD (see Fig. \ref{fig:PSDs}, in the electronic edition), consistent with their findings. However, the strength of this fluctuation above the continuum is not strong enough to suggest a QPO detection.

\begin{figure}[!t] \begin{center}
\includegraphics[width=0.95\columnwidth]{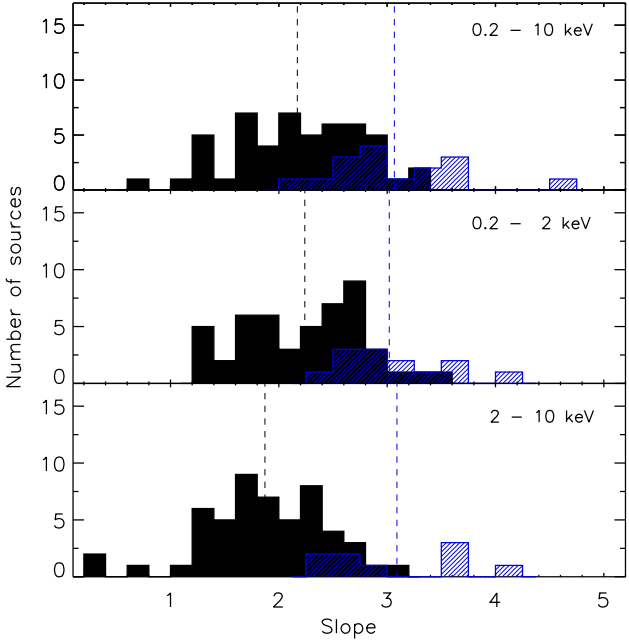}
\caption{Histograms of power-law slopes for Model A (black-filled histogram) and Model B (blue-dashed histogram). Top, middle, and bottom panels show the histograms for the total, soft, and hard bands, respectively. The dashed-lines show the mean value for each distribution. } 
\label{fig:histindex} 
\end{center} 
\end{figure}

	\subsection{Slopes}

\begin{figure}[!t] \begin{center}
\includegraphics[width=0.95\columnwidth]{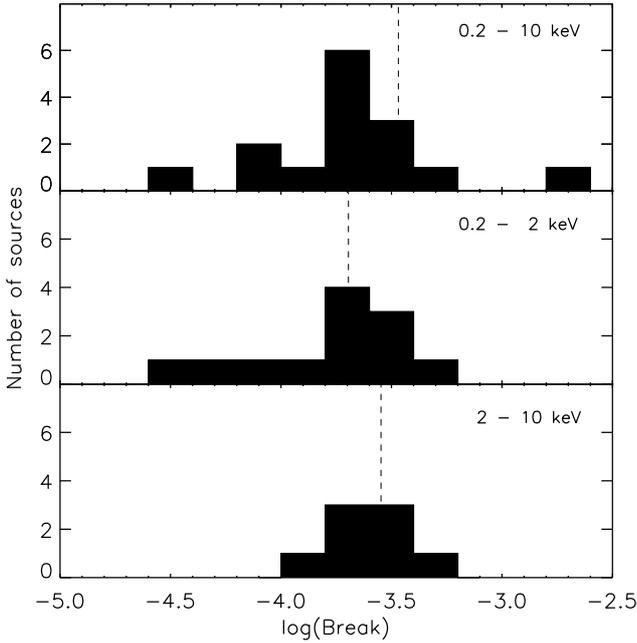}
\caption{Histograms of the bend frequency for Model B. Top, middle, and bottom panels show the histograms for the total, soft, and hard bands, respectively. The dashed-lines show the mean value for each distribution.} 
\label{fig:histbreak} 
\end{center} 
\end{figure}

The mean and standard deviation of the power law indices using Model A (for observation where Model B is not selected) was $\rm{\alpha =}$      2.01$\rm{\pm}$      0.01
 for the total energy band ($\rm{\alpha =}$      2.06$\rm{\pm}$      0.01
 and $\rm{\alpha =}$      1.77$\rm{\pm}$      0.01
 for the soft and hard bands, respectively). Fig. \ref{fig:histindex} (black filled histogram) shows the distribution of this parameter for the three energy bands\footnote{Where more than one observation was available for the same object we computed the weighted mean. This was calculated only in the case of multiple distinct observations; when multiple observations occurred close in time the periodogram data were fitted simultaneously (see Section \ref{sec:fit}).}. The distributions are  statistically indistinguishable according to the K-S test.

The mean value of the slope $\alpha$ when Model B model is preferred is $\rm{\alpha =}$      3.08$\rm{\pm}$      0.04
 for the total energy band ($\rm{\alpha =}$      3.03$\rm{\pm}$      0.01
 and $\rm{\alpha =}$      3.15$\rm{\pm}$      0.08
 for the soft and hard energy bands, respectively). These slopes are significantly steeper than those found from the Model A fit, i.e. where no PSD bend was detected. Fig. \ref{fig:histindex} (blue-dashed filled histogram) shows the distributions of slopes when Model B is preferred.

	\subsection{Frequency bend}

We have found PSD bends for $17$ AGN. Fig. \ref{fig:histbreak} shows the histograms of frequency bends. Note that the weighted mean value is reported when more than one observation is available for the same object.  The resulting mean values are $\rm{log(\nu_{br}) =     -3.47\pm      0.10}$
 for the total band (results for the soft and hard bands were almost identical).

Table \ref{tab:psd1} shows the PSD model parameters for the objects where Model B was selected. Of these, seven have not been previously reported in the literature: ESO\,113-G10, Mrk\,586, 1H0707-495, ESO\,434-G40, Circinus, NGC\,6860, and MS\,22549-3712  (although Mrk\,586 and MS\,22549-3712 are weak detections, see section \ref{sec:psdfit}). Table \ref{tab:frequencies} summarise the PSD information for the source bend detections in the present analysis together with a compilation of $17$ AGN with PSD bend frequencies from the literature. We have examined the \emph{XMM-Newton} data for all of them. This extend of the list of AGN with detected PSD bends to $24$ objects (or $22$ if we exclude Mrk\,586 and MS\,22549-3712). Appendix A gives a short description of the published bends together with a comparison with our results.

	\subsection{PSD bend amplitude}

\citet{Papadakis04} found that the PSD amplitude at the bend frequency, in terms of $\nu \times P(\nu)$, was roughly constant over a sample of AGN, at  $A_{psd} \sim 0.017$, based on an analysis of normalized excess variances. Table \ref{tab:frequencies} (Col. 11) shows $A_{psd}$ calculated from the estimated Model B parameters where a bend was detected. Although the mean value is $A = 0.009\pm 0.011$, roughly consistent with the \citet{Papadakis04} result, this is dominated by three AGN with high $A_{psd}$ values, namely 1H\,0707-495, NGC\,4051, and NGC\,4395 (excluding PKS\,0558-504 since it might not be a bend, see Section \ref{sec:psdfit}). These three sources are NLSy1s with low BH masses, which may indicate $A_{psd}$ is not constant over all AGN but depends on $M_{BH}$ and/or source type.

	\subsection{ Leakage distortion}\label{sec:leakage}

\begin{figure}[!t] \begin{center}
\includegraphics[width=1.\columnwidth]{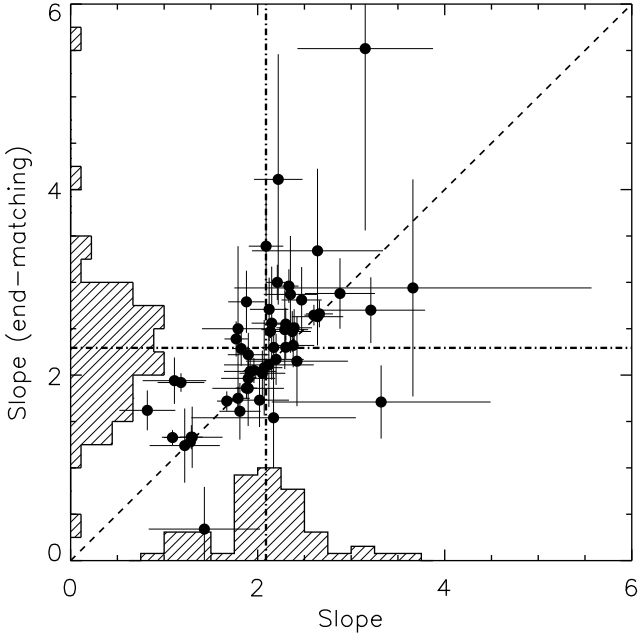}
\caption{Slopes obtained for Model A best-fit using the ``end-matching" routine versus those obtained without using ``end-matching". The dashed histograms are the normalized histograms for each distribution. The dashed line represents the one-to-one correlation and dot-dashed lines are the mean values for each distribution.}
\label{fig:leakage} 
\end{center} 
\end{figure}

A fraction of the power below and above the observed frequency range can `leak' into the observed bandpass due to the side-lobes of the Fejer kernel \citep[see][]{Priestley81, Percival93}. This is the problem of spectral leakage, discussed by \cite{Deeter82}, \cite{Uttley02} and \cite{Vaughan03B}. If the power spectrum is steep (e.g. $\alpha \gsim 2$) at or below the lowest observed frequency ($\nu_1 = 1/T$) this can lead to a significant distortion of the periodogram, as demonstrated by \cite{Uttley02}. However, the effect of leakage will be to reduce the sensitivity to bends, and QPOs, and bias the slope toward $\alpha = 2$, not to introduce spurious PSD features. Indeed, the detection of a bend to a flatter slope ($\alpha \lsim 2$) at low frequencies is an indication that leakage cannot be significant.  Some of the data sets for which the best-fitting model is a single power law (Model A) with slope $\alpha \approx 2$ are possibly be affected by leakage, for example NGC\,3516 and NGC\,4151 which are known to show bends to flatter slopes at frequencies lower than those covered by the \emph{XMM-Newton} observations \citep{Markowitz03}.  

\citet{Fougere85} discussed methods to recover accurate power spectral indicies in the presence of strong leakge. The simplest method that is reasonable effective at reducing leakage bias is 'end matching', removing a linear trend from the time series such that the first and last data points have equal values. 
We have used this technique for those sources best fitted to Model A for the 0.2--10 keV band. We recovered Model A as the preferred fit in all cases except IGR\,J21277+5656 ($\rm{log(\nu_{b}) =-3.84_{-0.22}^{+0.15}}$, $\rm{\alpha = 2.82_{-0.34}^{0.41}}$). A longer observation is needed to confirm the presence of the bend in this source. Moreover, after applying end-matching we are unable to constrain the parameters for Fairall\,9. (The time series of this time series is essentially a linear trend, no short timescale variability remains after end matching.)

Fig. \ref{fig:leakage} compares the slope estimates with and without end matching. Only PKS\,1547+79 shows a lower slope after the end-matching. However, the light-curve before end matching shows only a (roughly) linear trend; after end-matching no significant variability remains. The indices obtained after the end matching are on average higher than those obtained without end matching -- the mean index is $\rm{<\alpha>=2.29\pm 0.01}$ -- but still significantly lower than the mean high frequency index of those sources with detected power spectral bends.

% =============================================================================
% =============================================================================

\section{Discussion}\label{sec:discuss}

\subsection{Summary of results}

We have presented an analysis of the high frequency power density spectra (PSD) of a sample of $104$ AGN observed with  \emph{XMM-Newton}. These PSDs span up to $\approx 3$ orders of magnitude in frequency, ranging from minutes to over days. Our main results can be summarized as follows:

\begin{itemize}

\item Variability: $72$ per cent of the sample showed significant variability in at least one of the three energy bands tested. Moreover, most of LINERs in our sample do not vary (12 out of the 14 LINERs).

\item PSD shape: The PSDs of the majority of the variable AGN could be described by a single power-law (i.e. Model A) with a mean slope of $\alpha \approx 2$. In $17$ sources -- all Type-1 Seyferts (many are NLSy1s) but the Circinus galaxy -- we found a strong preference for the bending power-law model (i.e. Model B). For this subset, the slope above the bend frequency is steep, $\alpha \sim 3.0$, above a bend frequency typically around $\nu_{b} \sim 3.4\times 10^{-4}$ Hz. Spectral leakage, which can bias the slope estimates in some cases, does not appear to explain discrepancy between the slopes found for Model A and B.

\item QPO detection: None of the sources analyzed in our sample show a QPO, with the exception of KUG\,1031+398 (RE J$1034+396$), discussed extensively in the literature.

\end{itemize}

\subsection{Scaling relations}

\begin{figure}[!t] \begin{center}
\includegraphics[width=1.\columnwidth]{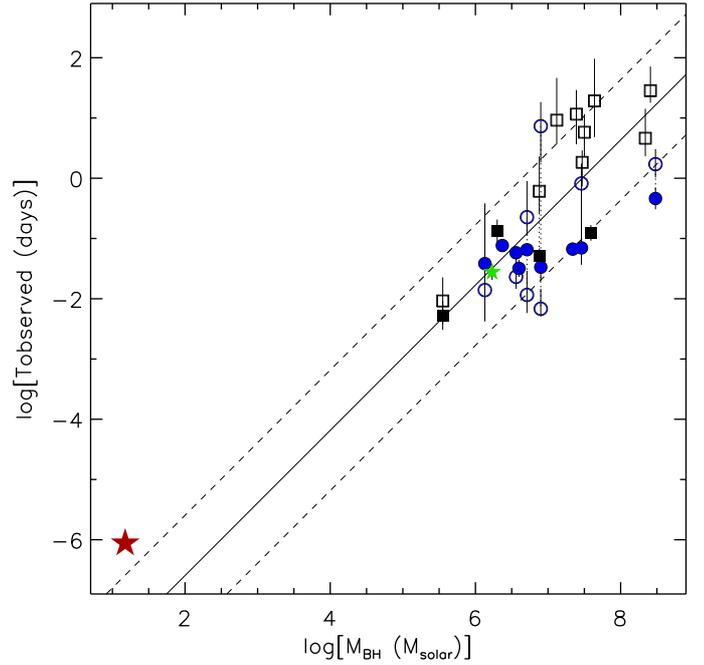} 
\caption{Observed bend timescale versus the BH mass. The continuous line is the best-fitting following eqn. \ref{eqn:mchardy2}. The dashed lines illustrate the $\pm 1$ dex region around this model. Circles represent NLSy1s, squares represent Type-1 Seyferts, and the small green star is the Type-2 Seyfert. The open symbols are data-points reported in the literature and filled symbols are the data-points reported here. The Cygnus X-1 data are shown as a red big star. A dotted line is used to link multiple frequency bends for the same object.}
\label{fig:Tobs_MBH} 
\end{center} 
\end{figure}

\begin{figure}[!t] 
\begin{center}
\includegraphics[width=1.\columnwidth]{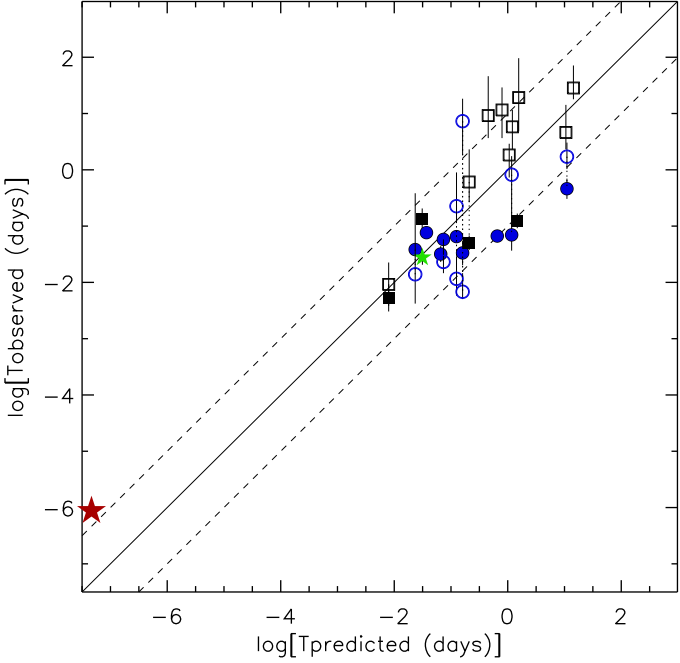} 
\caption{Observed bend timescale against the predicted value based on the best-fitting model following eqn. \ref{eqn:mchardy1}. Symbols are the same as explained in Fig. \ref{fig:Tobs_MBH}.}
\label{fig:Tbreak} 
\end{center} 
\end{figure}

\begin{figure}[!t] 
\begin{center}
\includegraphics[width=1.\columnwidth]{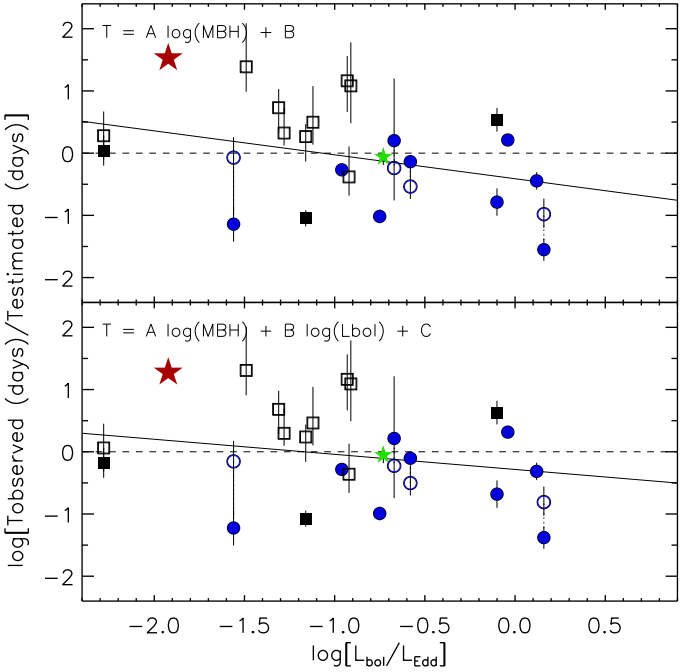} 
\caption{Ratio between the observed and predicted $T_b$ values against Eddington rate ($L_{bol}/L_{Edd}$) using eqn. \ref{eqn:mchardy2} (top) and eqn. \ref{eqn:mchardy1} (bottom). The continuous line shown in the top panel corresponds to the linear fit of the data. Symbols are the same as explained in Fig. \ref{fig:Tobs_MBH}.}
\label{fig:DifT_Ledd} 
\end{center} 
\end{figure}

Several papers have demonstrated a strong, approximately linear correlation between the PSD bend timescale ($T_{b} = 1/\nu_{b}$) and the black hole mass \citep[e.g.][]{Uttley02,Markowitz03,Vaughan05}, as expected from simple scaling arguments \citep{Shakura76, Fender06}. The simplest relation between timescale and fundamental AGN properties that we might consider is 
\begin{equation} \label{eqn:mchardy2} 
\log(T_{b}) = A \log(M_{BH}) + C 
\end{equation} 
where $A$ and $C$ are coefficients, and we expect $A \approx 1$ for a linear mass-timescale relation. \cite{McHardy06} presented an extension of this model allowing for dependence of $T_{b}$ on both $M_{BH}$ and bolometric luminosity $L_{bol}$: 
\begin{equation} \label{eqn:mchardy1} 
\log(T_{b}) = A \log(M_{BH}) + B \log(L_{bol}) + C 
\end{equation} 
Table \ref{tab:frequencies} gives values of $\log(\nu_b)$ for objects in which we find strong evidence for a bending power law PSD. From these we derive $T_{b}$ (in units of days). This table also includes values for $M_{BH}$ and $L_{bol}$ taken from the literature and the values of $\log(f_b)$, $M_{BH}$, and $L_{bol}$ compiled exclusively from previous literature. Together, these present the largest compilation of characteristic timescales for AGN to date. Note that the BH masses are derived by several different methods, but we have used estimates from reverberation mapping campaigns where these are available.

We fitted models following eqn. \ref{eqn:mchardy2} and \ref{eqn:mchardy1} to data from the $22$ AGN in table \ref{tab:frequencies}. We have used the $T_{b}$ values from our analysis when available and those from the literature otherwise, except for PKS\,558-504 for which we have used the value reported in the literature\footnote{Our bend in PKS\,558-504 could the result of a bump (or broad QPO) already reported by \citet{Papadakis10}.}. The models were fitted by simple linear regression\footnote{There are several reasons for not weighting the data according to the uncertainties on $T_{b}$. Firstly, the high frequency bend in X-ray binaries is not stationary but varies in frequency by a factor of $\sim$few, it is therefore plausible that the estimates of $T_{b}$ may not represent the `true' long-run average values if AGN power spectral are similarly non-stationary on very long timescales. Secondly, the $M_{BH}$ and $L_{bol}$ estimates are also subject to significant statistical and systematic errors that are difficult to quantify. These represent additional sources of variance in the data not accounted for by the confidence limits on $T_{b}$.} (i.e. unweighted least squares) on the log transformed variables, with $T_{b}$ in units of days, $L_{bol}$ in units of $10^{44}$ erg s$^{-1}$, and $M_{BH}$ in units of $10^6$ \Msun, as in \cite{McHardy06}.

Fitting equation \ref{eqn:mchardy2} gave parameter estimates $A = 1.09 \pm 0.21$ and $C = -1.70 \pm 0.29$, leaving a sum squared error (SSE) of $11.14$ (for $19$ degrees of freedom, dof). The data and best-fitting regression line are shown in Fig. \ref{fig:Tobs_MBH}. Fitting equation \ref{eqn:mchardy1} gave parameter estimates $A = 1.34 \pm 0.36$, $B = -0.24 \pm 0.28$, and $C = -1.88 \pm 0.36$, leaving a sum squared error (SSE) of $10.69$ (for $18$ dof). The parameter governing the luminosity dependence, $B$, is consistent with zero ($\rm{p = 0.38}$ for 18 dof). Fig. \ref{fig:Tbreak} compares the bend timescale predicted by this relation (based on the observed $M_{BH}$ and $L_{bol}$) to the observed $T_{b}$ values \citep[compare with Fig. 2 of][]{McHardy06}. Fig. \ref{fig:DifT_Ledd} shows the residuals (as a ratio of observed to model $T_{b}$) against $L_{bol} / L_{Edd}$ for the two different models. 

In order to test the reliability of these parameters we have used the `bootstrap' method \citep{Efron93}. We generated $N=10^4$ simulated datasets, each comprising $22$ 'observations' (of $T_{b}$, $M_{BH}$, $L_{bol}$) generated by drawing an AGN at random (with replacement) from the real data. This provides a simple alternative assessment of the parameter confidence intervals. The $68.3$ per cent confidence intervals on the parameters are: $A = [1.02,  1.66]$, $B = [-0.52, -0.046]$ and $C = [-2.12, -1.65]$. The boostrap confidence intervals are slightly narrower than the standard regression intervals.

In order to test how well these scaling relations work over the full range of black hole masses we also show representatives values for the BH-XRB Cygnus X-1 (red star in the figures)\footnote{We used the black hole mass of $M_{BH} = 15 \pm 1 \rm{M_{\odot}}$ recently presented by \citet{Orosz11}. For the bolometric luminosity and characteristic time scale we took the average of several estimates of the PSD bend frequency from \citet{Axelsson06} and bolometric flux from \citet{Wilms06} (using the data from their Model 5, Table 1 -- see \citet{McHardy06} for justification of this choice of model). The observations were chosen to be those with simultaneous bend frequency and bolometric flux estimates. The luminosity $L_{bol}$ was derived assuming a distance of $\rm{D = 1.86\pm 0.12}$ kpc \citep{Reid11}. The final bolometric luminosity and frequency bend are $L_{bol}= 2.26 \pm 0.73 \times 10^{37}$ erg s$^{-1}$ and $\nu_{b}= 13.2 \pm 6.0$ Hz, respectively.}. The Cygnus X-1 points were not included in the fitting, yet are clearly consistent with an extrapolation to much lower $M_{BH}$, strongly supporting the reliability of such relations over the full range of $M_{BH}$. Indeed, fitting the two models including the Cygnus X-1 data resulted in parameter estimates consistent with those given above.

The main difference compared with the results obtained by \citet{McHardy06} is a weak dependence of $T_{b}$ on $L_{bol}$ in the present analysis. This remains the case when the fitting is repeated with or without the Cygnus X-1 data ($\rm{B=-0.27\pm0.27}$), using a lower mass estimate for NGC\,4395\footnote{NGC\,4395 is the object in our sample with the most `leverage' on the regression model. It could have a lower black hole mass as discussed by \citealt{Vaughan05} and \citealt{Uttley05} than the reverberation mapping mass of \citet{Peterson05}.}, or using a weighted least squares regression (i.e. making use of the confidence intervals on $T_{b}$). However, if we use smaller black-hole masses for NGC\,4395 \citep[($\rm{log(M_{BH})=4.5}$, see][]{Vaughan05,Uttley05} and NGC\,5506 \citep[$\rm{log(M_{BH})=6.5}$, see][]{McHardy06}, and NGC\,6860 is removed\footnote{We tried to remove NGC\,6860 because it is one of the drop-outs in our correlation.}, the dependence with the $L_{bol}$ found by \citet{McHardy06} is recovered ($\rm{B=-0.70\pm0.30}$, $\rm{p=0.01}$). Thus, this weak dependence on $L_{bol}$ could be due to either the fact that our sample is more complete sample (more objects and new estimates from the old bends) or due to uncertainties on the BH mass and/or $\rm{L_{bol}}$ estimates. A bigger sample with better estimates on the BH mass and $\rm{L_{bol}}$ is need to check the dependence on $\rm{L_{bol}}$.

\subsubsection{BLR versus variability}

\begin{figure}[!t] 
\begin{center}
\includegraphics[width=1.\columnwidth]{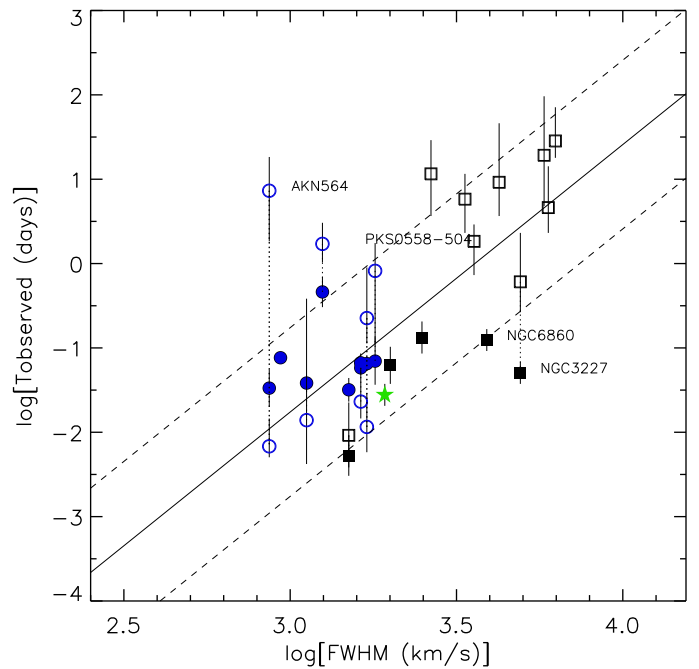} 
\caption{Observed characteristic timescale versus the FWHM of H$\rm{_{\beta}}$ expressed in km$\rm{~s^{-1}}$. Symbols are the same as explained in Fig. \ref{fig:Tobs_MBH}.} 
\label{fig:Tfwhm} 
\end{center} 
\end{figure}

\cite{McHardy06} also presented strong correlation between $T_{b}$ and the width of the (permitted) optical lines, $V$, specifically H$\beta$. We have compiled optical line widths for the objects in Tables \ref{tab:psd1} and \ref{tab:frequencies} based on the available literature. For most sources we have used as $V$ the full width at half maximum (FWHM) of the H$\beta$ line, the exceptions are NGC\,5506 and ESO\,434-G40, which are both heavily obscured and for which we have used the FWHM of the Pa$\beta$ line. Fig. \ref{fig:Tfwhm} shows the resulting timescale-FWHM relation. The linear correlation coefficient is $r = 0.692$ (highly significant; $p=5.0\times 10^{-4}$). Simple linear regression of $\log (T_{b})$ on $\log(V)$ was used to fit a relation of the form 
\begin{equation} \label{eqn:Tfwhm} 
\log(T_{b}) =   D \log(V) + E. 
\end{equation} 
The best-fitting parameter values were $D = 2.9  \pm  0.7 $ and $E =  -10.2 \pm 2.3$, leaving a sum squared error (SSE) of $13.47$ (for $19$ degrees of freedom, dof). This relation is again slightly flatter than the $D=4.2$ found by  \cite{McHardy06}. The $68.3$ per cent confidence intervals on these parameters calculated by the bootstrap method are  $D = [2.2,  3.5]$ and $E = [-12.3,  -7.9]$. % Again, using $T_{b}$ estimates from the literature values in preference to those presented in this paper, and the lower $M_{BH}$ estimate for NGC\,4395, gave parameters in closer agreement with those of \cite{McHardy06}.

\subsection{PSD shapes} 

For the ten objects with well-constrained Model B parameters, the high frequency slope is steep ($\alpha \approx 3$). Such steep high frequency PSDs have been reported in the literature \citep[e.g. Mrk\,335 and Mrk\,766 by][respectively, see our Table \ref{tab:frequencies} for details]{Arevalo08,Vaughan03C}, but with little explicit discussion of this point.

There seems to be no fundamental physical requirement for the PSD slope to be steeper than $\alpha = 2$ at high frequencies \citep[see footnote $7$ of][]{Vaughan11}. Indeed, $\alpha \approx 2$ seems to be the norm for BH-XRBs in `hard' states in the $\sim 5-50$ Hz regime, above which there may be further steeping \citep{Nowak99, Revnivtsev00}. The `soft' states may show steeper PSDs \citep{Revnivtsev00,McHardy04}, and so the steep high frequency PSDs observed from these AGN may support the argument that AGN are supermassive analogues of soft state BH-XRBs. There are alternative explanations that must be considered, however. These include the fact that the AGN PSDs presented here are based on \emph{XMM-Newton} data with a relatively soft X-ray response, while the BH-XRB results are based on observations at harder X-rays with \emph{RXTE}, and therefore the known energy dependence of the high frequency slope \citep{Nowak99,Vaughan03B,McHardy04} may explain the apparently steeper slopes in AGN. (We have relatively little information about the PSDs of BH-XRBs at $\sim 1$ keV. See \citealt{Uttley11} for a rare example based on a bright `hard' state observation.) Another possibility is that the steeper slopes are the result of a selection effect and are not representative of the larger population of bright AGN. The PSD bend is easier to detect when the high frequency slope is steep, and when the bend is at relatively low frequency. We cannot exclude the possibility that only a subset of AGN show such steep slopes but these are the ones for which the PSD bend, and hence high frequency slope, is easiest to constrain.

The majority of the sample did not show evidence of a bend (i.e. insufficient evidence to select Model B over Model A in a LRT). The mean value of the PSD index for these objects was $\alpha \approx 2$. The simplest interpretation is that these generally more massive objects have lower $\nu_{b}$, and hence we see only the high frequency part of the PSD. This slope is $\alpha \approx 2.3$ when the leakage distortion is reduced using end matching, but this remains flatter than the high frequency slope for objects that do show evidence for a PSD bend. It remains possible that a more rigorous method to account for the leakage bias \citep[e.g. the methodology described by][]{Uttley02} could recover steeper slopes. But this could also result from a selection effect. If AGN show a wide range of high frequency PSD slopes, e.g. $\alpha \sim 2-4$ above the bend frequency, then those with flatter slopes will be more difficult to detect, as the change in slope (from the low frequency index of $1$) is less pronounced.  

In order to study whether we detect all the expected bends, we have computed the predicted characteristic time scale $T_b$ for sample members with $M_{BH}$ estimates ($70$ out of the $104$), assuming bolometric luminosity of $L_{bol} = 30 \times L_{2-10}$ (estimated as explained in Section \ref{sec:lum}). The PSD bend timescales is predicted to be within the observed frequency range for only $17$ sources. Of these we do indeed detect bends in $13$. The exceptions are 1H\,419-577, NGC\,4593, IRAS\,13224-3809, and NGC\,7314. However, note that if we use the relation found by \citet{McHardy06} IRAS\,13224-3809 and NGC\,7314 are not expected to show a bend in our frequency range. For the rest of the sample members the predicted $T_{b}$ value places it outside of the observed PSD bandpass. Moreover, all the non-variable sources have an expected bend well above the range analyzed here. 

\subsection{Lack of QPOs}

We find no strong evidence for highly coherent oscillations (QPOs) beyond the well known example of KUG\,1031+398. Some authors have pointed to the extremely soft spectrum of this source as an explanation of why this source appears to be unique in showing a QPO. In fact, even for this source the QPO appears to be transient \citep{Middleton09,Middleton11}. Less coherent oscillations, i.e. QPOs with a broader features in the PSD, are common to BH-XRBs, especially in harder spectral states, but are difficult to detect in the majority of the currently available data, as discussed by \citet{Vaughan05B} and \citet{Vaughan11}.

% =============================================================================
% =============================================================================

\begin{acknowledgements} 

This research has made use of the NASA Astrophysics Data System Bibliographic Services, and the NASA/IPAC Extragalactic Data base (NED) which is operated by the Jet Propulsion Laboratory, California Institute of Technology, under contract with the National Aeronautics and Space Administration. This paper is based on observations obtained with \emph{XMM-Newton}, an ESA science mission with instruments and contributions directly funded by ESA Member States and the USA (NASA). OGM acknoledges to the C5D00070-2006 grant of the MICINN for the finantial support. OGM thanks Dr. Iossif Papadakis for his help understanding variability processes in AGN and the interpretation of the results of the present study. 

\end{acknowledgements}

% =============================================================================
% =============================================================================

% =============================================================================
% =============================================================================

\onecolumn
\scriptsize{
\begin{longtable}{l l c c c c c r c c c }
\hline\hline                 % inserts double horizontal lines
Num & Name        & Type  & z  &  $log(M_{BH})$ &  $log(L_{bol})$ & Ref. & ObsID & Expos. & L(2-10 keV) &	 Var. \\ % table heading 
(1) &      (2)       &  (3)   &   (4) & (5)&  (6)  &   (7)  &    (8)  & (9)  & (10) &  (11)  \\ \hline             	 % inserts single horizontal line
    &                &        &       &    &       &        &       &      &    &     \\
\endfirsthead
\hline\hline                 % inserts double horizontal lines
Num & Name        & Type  & z  &  $log(M_{BH})$ &  $log(L_{bol})$ & Ref. & ObsID & Expos. & L(2-10 keV) &	 Var. \\ % table heading 
(1) &      (2)       &  (3)   &   (4) & (5)&  (6)  &   (7)  &    (8)  & (9)  & (10) &  (11)  \\ \hline             	 % inserts single horizontal line
    &                &        &       &    &       &        &       &      &    &     \\
\endhead
    1&         Mrk335&  NLSY1&  0.0258&       7.3&      44.7& (1)&   306870101&   114.5&    43.4&   (TSH) \\
     &               &       &        &          &          &    &   600540501&    80.5&    42.9&   (TSH) \\
     &               &       &        &          &          &    &        0601&   130.1&    42.9&   (TSH) \\
    2&           IZw1&  NLSY1&  0.0611&       7.4&      45.1& (2)&   300470101&    79.8&    43.6&   (TSH) \\
    3&         NGC315&     L2&  0.0149&       7.2&      43.3& (3)&   305290201&    46.6&    41.9&   (NNN) \\
    4&      PGC643840&   S1.0&  0.1650&       8.9&      45.8& (2)&   205680101&    97.9&    44.3&   (NNN) \\
     &               &       &        &          &          &    &   401930101&    41.8&    44.3&   (NNN) \\
    5&     ESO113-G10&   S1.8&  0.0260&          &      44.2& (2)&   301890101&    94.8&    42.7&   (TSH) \\
\end{longtable}
}

\begin{center} 
\scriptsize{  
 \begin{longtable}{l r c c c c c c }  
\hline\hline                 % inserts double horizontal lines  
 Name		& ObsID  &  \multicolumn{2}{c}{0.2-10 keV}    &  \multicolumn{2}{c}{0.2-2 keV}  &  \multicolumn{2}{c}{2-10 keV}      \\  % table heading 
 		&               &  log(N)  & $\alpha$       &  log(N)  & $\alpha$      &  log(N)  & $\alpha$       \\ % table heading   
     (1)	&  (2)   &   (3)       & (4)  &   (5)       & (6)  &   (7)       & (8)	  \\ \hline		    % inserts single horizontal line  
		&	 &	       &	 &	       &	 &	       &	       	  \\  
\endfirsthead  
\hline\hline                 % inserts double horizontal lines  
 Name		& ObsID  &  \multicolumn{2}{c}{0.2-10 keV}    &  \multicolumn{2}{c}{0.2-2 keV}  &  \multicolumn{2}{c}{2-10 keV}      \\ % table heading  
 		&               &  log(N)  & $\alpha$       &  log(N)  & $\alpha$      &  log(N)  & $\alpha$       \\ % table heading   
     (1)	&  (2)   &   (3)       & (4)  &   (5)       & (6)  &   (7)       & (8)	  \\ \hline		    % inserts single horizontal line  
		&	 &	       &	 &	       &	 &	       &	       	  \\  
\endhead  
         Mrk335 &  306870101 &   -9.1 $\rm{_{   -0.5}^{+    0.5}}$ &     2.7 $\rm{_{   -0.1}^{+    0.2}}$&   -8.9 $\rm{_{   -0.5}^{+    0.5}}$ &     2.6 $\rm{_{   -0.1}^{+    0.2}}$&   -6.2 $\rm{_{   -0.6}^{+    0.6}}$ &     2.0 $\rm{_{   -0.2}^{+    0.2}}$ \\
                &  600540501 &   -7.8 $\rm{_{   -0.4}^{+    0.4}}$ &     2.4 $\rm{_{   -0.1}^{+    0.1}}$&   -7.6 $\rm{_{   -0.4}^{+    0.4}}$ &     2.3 $\rm{_{   -0.1}^{+    0.1}}$&   -6.2 $\rm{_{   -0.8}^{+    0.6}}$ &     2.0 $\rm{_{   -0.2}^{+    0.2}}$ \\
                &       0601 &                                     &                                     &                                     &                                     &                                     &                                     \\
           IZw1 &  300470101 &   -8.2 $\rm{_{   -0.9}^{+    0.8}}$ &     2.4 $\rm{_{   -0.2}^{+    0.2}}$&   -8.2 $\rm{_{   -0.9}^{+    0.8}}$ &     2.4 $\rm{_{   -0.2}^{+    0.2}}$&   -5.4 $\rm{_{   -1.3}^{+    1.1}}$ &     1.7 $\rm{_{   -0.3}^{+    0.3}}$ \\
     ESO113-G10 &  301890101 &   -6.6 $\rm{_{   -0.6}^{+    0.5}}$ &     2.0 $\rm{_{   -0.2}^{+    0.2}}$&   -6.4 $\rm{_{   -0.6}^{+    0.5}}$ &     2.0 $\rm{_{   -0.2}^{+    0.2}}$&   -5.1 $\rm{_{   -0.9}^{+    0.8}}$ &     1.7 $\rm{_{   -0.2}^{+    0.2}}$ \\
       Fairall9 &  605800401 &   -7.8 $\rm{_{   -0.8}^{+    0.8}}$ &     2.1 $\rm{_{   -0.2}^{+    0.2}}$&   -7.8 $\rm{_{   -0.8}^{+    0.8}}$ &     2.1 $\rm{_{   -0.2}^{+    0.2}}$&   -7.8 $\rm{_{   -1.5}^{+    1.2}}$ &     2.0 $\rm{_{   -0.3}^{+    0.3}}$ \\
        NGC526A &  150940101 &  -13.4 $\rm{_{\clubsuit}^{+    4.0}}$ &     3.3 $\rm{_{   -1.0}^{+    1.3}}$&  -13.9 $\rm{^{\clubsuit      }}$ &     3.5 $\rm{_{   -1.2}^{+    1.7}}$&  -11.8 $\rm{_{\clubsuit}^{+    3.5}}$ &     3.0 $\rm{_{   -0.8}^{+    1.2}}$ \\
\end{longtable} 
} 
\end{center} 

 \begin{center}
 \scriptsize{
 \begin{longtable}{l r c c c c c c c c c c c c}
\hline\hline  	       % inserts double horizontal lines
 Name 	  & ObsID  &  \multicolumn{4}{c}{0.2-10 keV}  &  \multicolumn{4}{c}{0.2-2 keV} &  \multicolumn{4}{c}{2-10 keV} \\ \cline{4-5}	\cline{8-9}  \cline{12-13}  % table heading 
		  &	   &		 &	  &		  &	&	 &	  &		  &    &	 &	  &		  &		    \\
		  &	   &  log(N)  & $\alpha$  &  log($\rm{\nu_{br}}$) &	log(p)  &  log(N)  & $\alpha$  &  log($\rm{\nu_{br}}$) &    log(p)  &  log(N)  & $\alpha$  &  log($\rm{\nu_{br}}$) &  log(p) \\ % table heading 
     (1)	  &  (2)   &   (3)	 & (4)    &  (5)	  &   (6)    &   (7)	   & (8)  &  (9)	  &   (10)  &	(11)	   & (12) &  (13)	  &   (14) \\ \hline		      % inserts single horizontal line
		  &	   &		 &	  &		  &	&	 &	  &		  &    &	 &	  &		  &		    \\
\endfirsthead
\hline\hline  	       % inserts double horizontal lines
 Name 	  & ObsID  &  \multicolumn{4}{c}{0.2-10 keV}  &  \multicolumn{4}{c}{0.2-2 keV} &  \multicolumn{4}{c}{2-10 keV} \\ \cline{4-5}	\cline{8-9}  \cline{12-13}  % table heading 
		  &	   &		 &	  &		  &	&	 &	  &		  &    &	 &	  &		  &		    \\
		  &	   &  log(N)  & $\alpha$  &  log($\rm{\nu_{br}}$) &   log(p)	&  log(N)  & $\alpha$  &  log($\rm{\nu_{br}}$) &   log(p)   &  log(N)  & $\alpha$  &  log($\rm{\nu_{br}}$) &  log(p) \\ % table heading 
     (1)	  &  (2)   &   (3)	 & (4)    &  (5)	  &   (6)    &   (7)	   & (8)  &  (9)	  &   (10)  &	(11)	   & (12) &  (13)	  &   (14) \\ \hline		      % inserts single horizontal line
		  &	   &		 &	  &		  &	&	 &	  &		  &    &	 &	  &		  &		    \\
\endhead
         Mrk335 &  306870101 &   -2.3 $\rm{_{   -0.1}^{+    0.2}}$ &   3.7 $\rm{_{   -0.4}^{+    0.4}}$ &  -3.76 $\rm{_{  -0.12}^{+   0.10}}$ &       -6.3 &   -2.3 $\rm{_{   -0.1}^{+    0.2}}$ &   3.6 $\rm{_{   -0.3}^{+    0.4}}$ &  -3.80 $\rm{_{  -0.13}^{+   0.10}}$ &       -5.4 &   -2.3 $\rm{_{   -0.1}^{+    0.1}}$ &   3.7 $\rm{_{   -0.6}^{+    1.0}}$ &  -3.64 $\rm{_{  -0.14}^{+   0.10}}$ &       -4.0 \\
     ESO113-G10 &  301890101 &   -2.2 $\rm{_{   -0.2}^{+    0.2}}$ &   3.0 $\rm{_{   -0.4}^{+    0.6}}$ &  -3.73 $\rm{_{  -0.26}^{+   0.18}}$ &       -2.3 &   -2.2 $\rm{_{   -0.2}^{+    0.2}}$ &   2.9 $\rm{_{   -0.4}^{+    0.6}}$ &  -3.75 $\rm{_{  -0.30}^{+   0.19}}$ &       -2.0 &   -2.2 $\rm{_{   -0.1}^{+    0.2}}$ &   4.5 $\rm{_{   -1.7}^{+    3.8}}$ &  -3.64 $\rm{_{  -0.18}^{+   0.12}}$ &       -1.8 \\
         Mrk586 &  048740101 &   -2.2 $\rm{_{   -0.2}^{+    0.3}}$ &   4.2 $\rm{_{   -0.8}^{+    1.3}}$ &  -3.92 $\rm{_{  -0.22}^{+   0.15}}$ &       -1.7 &   -2.2 $\rm{_{   -0.2}^{+    0.4}}$ &   3.9 $\rm{_{   -0.8}^{+    1.2}}$ &  -3.94 $\rm{_{  -0.26}^{+   0.17}}$ &       -1.5 &   -1.9 $\rm{_{   -0.2}^{+    0.2}}$ &  10.7 $\rm{^{\clubsuit      }}$ &  -3.78 $\rm{_{  -0.09}^{+   0.11}}$ &       -2.1 \\
    PKS0558-504 &  555170201 &   -1.5 $\rm{_{   -0.2}^{+    0.2}}$ &   2.5 $\rm{_{   -0.1}^{+    0.1}}$ &  -4.60 $\rm{_{  -0.20}^{+   0.15}}$ &       -2.5 &   -1.5 $\rm{_{   -0.2}^{+    0.2}}$ &   2.5 $\rm{_{   -0.1}^{+    0.1}}$ &  -4.59 $\rm{_{  -0.20}^{+   0.15}}$ &       -2.6 &   -1.3 $\rm{_{   -0.2}^{+    0.4}}$ &   2.4 $\rm{_{   -0.2}^{+    0.2}}$ &  -4.78 $\rm{_{  -0.47}^{+   0.26}}$ &       -1.0 \\
                &       0301 &                      &                      &                      &         \\
                &       0401 &                      &                      &                      &         \\
                &       0501 &                      &                      &                      &         \\
\end{longtable}
}
\end{center}

\newpage
\scriptsize{
\begin{longtable}{l c c c c c c c c c c}
\hline\hline                 % inserts double horizontal lines
    &          &        &             &      &	  &		     	 &	       &  &  \\
\multicolumn{11}{c}{Breaks reported in this study} \\  \cline{4-8}
    &          &        &             &      &	  &		     	 &	       &  &  \\
 Name	     & Ref.& Type & Satellite &  Band &  log($\rm{M_{BH}}$) 	& $\rm{L_{bol}}$  & FWHM($\rm{H\beta}$)   &$\rm{\alpha}$    &  log($\rm{\nu_{br}}$) & $\rm{\nu_{br}\times P(\nu_{br})}$ \\ % table heading
	     &     &      &          &       &       			&        &     (km/s)  &             &			   &			($\rm{\times 10^{-3}}$)    \\
   (1)	     &  (2)& (3)  &  (4)     &   (5) &   (6) 			&   (7)  &(8)              &   (9)		   &	(10)		& (11)   \\ 	 \hline	% inserts single horizontal line
\endfirsthead
 Name	     & Ref.& Type & Satellite &  Band &  log($\rm{M_{BH}}$) 	& $\rm{L_{bol}}$    & FWHM($\rm{H\beta}$)  &$\rm{\alpha}$    &  log($\rm{\nu_{br}}$) & $\rm{\nu_{br}\times P(\nu_{br})}$ \\ % table heading 
	     &     &      &          &       &       			&        &     (km/s)   &         &			   &			($\rm{\times 10^{-3}}$)    \\
   (1)	     &  (2)& (3)  &  (4)     &   (5) &   (6) 			&   (7)  &(8)              &   (9)		   &	(10)	& (11)	   \\ 	 \hline	% inserts single horizontal line
\endhead
         Mrk335 & 1 &      NLSY1 & XMM &   0.2-10 &    7.34(1) &   44.69(1) &      1629 &     3.73$\rm{_{   -0.15}^{+    0.15}}$ &    -3.76$\rm{_{   -0.11}^{+    0.11 }}$ &      2.2 $\rm{\pm}$      0.3\\
     ESO113-G10 & 1 &         S1 & XMM &   0.2-10 &            &   44.17(7) &           &     2.98$\rm{_{   -0.22}^{+    0.22}}$ &    -3.73$\rm{_{   -0.22}^{+    0.22 }}$ &      2.9 $\rm{\pm}$      0.5\\
    PKS0558-504 & 1 &      NLSY1 & XMM &   0.2-10 &   8.48(12) &   46.74(6) &      1250*&     2.52$\rm{_{   -0.19}^{+    0.19}}$ &    -4.60$\rm{_{   -0.18}^{+    0.18 }}$ &     16.6 $\rm{\pm}$      3.1\\
     1H0707-495 & 1 &      NLSY1 & XMM &   0.2-10 &    6.37(2) &   44.43(2) &       935 &     2.50$\rm{_{   -0.25}^{+    0.25}}$ &    -3.82$\rm{_{   -0.07}^{+    0.07 }}$ &     24.2 $\rm{\pm}$      1.4\\
     ESO434-G40 & 1 &         S1 & XMM &   0.2-10 &    6.30(3) &   44.30(3) &      2490*&     3.52$\rm{_{   -0.38}^{+    0.38}}$ &    -4.06$\rm{_{   -0.19}^{+    0.19 }}$ &      1.0 $\rm{\pm}$      0.2\\
        NGC3227 & 1 &         S1 & XMM &   0.2-10 &    6.88(4) &   43.86(1) &      4920 &     3.44$\rm{_{   -0.20}^{+    0.20}}$ &    -3.64$\rm{_{   -0.13}^{+    0.13 }}$ &      2.2 $\rm{\pm}$      0.3\\
    KUG1031+398 & 1 &      NLSY1 & XMM &   0.2-10 &    6.60(5) &   44.82(4) &      1500 &     3.26$\rm{_{   -0.09}^{+    0.09}}$ &    -3.44$\rm{_{   -0.14}^{+    0.14 }}$ &      1.1 $\rm{\pm}$      0.1\\
        NGC4051 & 1 &      NLSY1 & XMM &   0.2-10 &    6.13(4) &   43.56(1) &      1120 &     2.63$\rm{_{   -0.38}^{+    0.38}}$ &    -3.52$\rm{_{   -0.09}^{+    0.09 }}$ &     25.3 $\rm{\pm}$      1.9\\
\end{longtable} 	
}

\twocolumn
% ---------------------------------------------
% =============================================================================
\normalsize
\begin{appendix}

\section*{Appendix A: PSD breaks/bends in the literature}

Here we  give a short summary of previously reported  PSD frequency bends/breaks together  and a comparison with our results. Table \ref{tab:frequencies} shows the parameter values discussed here.  It should be noted that the {\it XMM-Newton} data discussed in the present paper are sensitive to bends only in the frequency range from $\gsim 10^{-4}$ to $\sim$few$\times 10^{-3}$.

$\rm{\bullet}$ \emph{Mrk\,335}: The 0.2-10 keV band PSD of Mrk\,335 was reported by \citet{Arevalo08} using the same data-set used here. Our results are   consistent with theirs.

$\rm{\bullet}$ \emph{NGC\,3227}:  \citet{Uttley05} analyzed the 2-10 keV band PSD of NGC\,3227 using \emph{RXTE} data and reported  a break at $2.6 \times 10^{-5}$ Hz, a factor $\sim 9$ lower than our result. \citet{Kelly11} analysed the \emph{RXTE} and \emph{XMM-Newton} data using a different method for power spectrum estimation, and obtained a bend frequency of $\sim 3.7 \times 10^{-5}$ Hz. Some of the discrepancy between these results and ours may be due to the different energy bands and responses used in the different the analyses, i.e. 2--10 keV \emph{RXTE} data used by \citet{Uttley05} compared to 0.2--10 keV \emph{XMM-Newton} data used here, although it seems unlikely this can account for all the difference.

$\rm{\bullet}$ \emph{KUG\,1031+398}: V10  presented an analysis of the KUG\,1031+398 data using essentially the same method as in the present analysis. The two sets of results are fully consistent with each other.

$\rm{\bullet}$ \emph{NGC\,3516}:  \citet{Markowitz03} reported the 2-10 keV PSD of NGC\,3516 using \emph{RXTE} data. They found a high-frequency break at $\rm{log(\nu_{b})= -5.7\pm0.4}$  (slopes of 1 and 2 below and above the frequency break, respectively). These are outside the frequency range available to the \emph{XMM-Newton} data. As expected we find no strong evidence for a bend within the frequency bandpass considered here, and the Model A slope is consistent with the high frequency slope of \citet{Markowitz03}.

$\rm{\bullet}$ \emph{NGC\,3783}:  The power spectrum of this objects has been discussed by \citet{Markowitz03}, \citet{Markowitz05}, \citet{Summons07}, \citet{Arevalo09} and \citet{Kelly11} using combinations of \emph{RXTE} and \emph{XMM-Newton} data. These papers report a single bend in the PSD at $\sim 6.2 \times 10^{-6}$Hz, outside the frequency bandpass available to the \emph{XMM-Newton} data. The power spectral slopes reported here are slightly flatter than those reported by \citet{Markowitz05}.

$\rm{\bullet}$ \emph{NGC\,4051}:  The PSD of NGC\,4051 has discussed by \citet{McHardy04}, \citet{Miller10} and \citet{Vaughan11}. \citet{McHardy04} used \emph{RXTE} and \emph{XMM-Newton} data and found a frequency bend at $\rm{log(\nu_{b}) \sim -3.1}$, albeit with a large uncertainty. \citet{Vaughan11} analysed a series of \emph{XMM-Newton} observations, finding broadly similar results but a slightly lower bend frequency ($\log(\nu_{b}) \sim -3.7$). Here we report a PSD bend at $\rm{log(\nu_{b}) \sim -3.5}$,  consistent with those reported by with other analyses.

$\rm{\bullet}$ \emph{NGC\,4151}: \citet{Markowitz03}  discussed the 2-10 keV PSD of NGC\,4151 using \emph{RXTE} data, and reported a break frequency (using a broken power law model) of $\sim 10^{-6}$ Hz, outside of the frequency bandpass available to the \emph{XMM-Newton} data. The single power-law model (Model A) applied to the {\it XMM-Newton} data gave a best-fitting slope ($2.1$) identical to the high frequency slope reported by \citet{Markowitz03}.

$\rm{\bullet}$ \emph{Mrk\,766}:   The PSD of Mrk\,766 was studied by \citet{Vaughan03C}, \citet{Markowitz07} and \citet{Kelly11}. The main parameters (bend frequency and high frequency slope) are broadly compatible with our findings. The bend frequency reported by \citet{Markowitz07} is slightly higher, but we assumed a low frequency slope of $1$ whereas they fitted for this parameter and found a slightly steeper slope.

$\rm{\bullet}$ \emph{NGC\,4395}:  \citet{Vaughan05} report on the PSD using \emph{XMM-Newton} data, and their results are completely consistent with those presented here (their fitting method differs from that used presently). \citet{Shih03} previously used \emph{ASCA} data to study the PSD of NGC\,4395, finding a PSD break at $\log(\nu_{b})= -3.5$ (with slopes of $\alpha \sim 1$ and $\alpha \sim 1.8$  at low and high frequencies, respectively). However, the PSD reported by \citet{Shih03} lacked data in the frequency range $\log(\nu) = [-4, -3.4]$, which hampered accurate estimation of the frequency break.

$\rm{\bullet}$ \emph{Fairall 9}:  \citet{Markowitz03} and \citet{Kelly11} discuss the PSD of this source using \emph{RXTE} data. The former paper reports an upper limit on the bend frequency of $\log(\nu_{b}) \sim -6.4$, and the latter paper gives an estimate of $\log(\nu_{b}) \approx -5.6$, outside of the frequency bandpass available to the \emph{XMM-Newton} data; as expected given these results we find no strong evidence for a bend in the present analysis. The single power-law fit (i.e. Model A) shows a slope consistent with the previous high frequency slope estimates.

$\rm{\bullet}$ \emph{MCG$-$06$-$30$-$15}:  \citet{Vaughan03B} and \citet{McHardy05} have reported on the PSD of this source. In the present paper we report and bend frequency higher than that reported by \citet{Vaughan03B}. However, they used a sharply broken power law while here we use a bending power law model, which means the frequency parameter is not strictly the same. \citet{McHardy05} analysed together the \emph{RXTE} and \emph{XMM-Newton} data and found a good fit using the bending power law model. Their estimated bend frequency is a factor $\approx 3$ lower than that reported in the present paper. This may in part be due to the use of different energy bands between the analyses (e.g. 4--10 keV for \emph{RXTE} and 0.2--10 keV \emph{XMM-Newton}).

$\rm{\bullet}$ \emph{IC\,4329A}:  \citet{Markowitz09}  reported the 3-10 keV band PSD. They fitted the PSD  to a broken power law with a break at $\rm{log(\nu_{b})\sim -5.6}$, outside of the frequency bandpass avilable to the \emph{XMM-Newton} data. They report a high frequency slope $\alpha = 2.3_{-0.4}^{+0.8}$ consistent with the slope using with Model A in the present study.

$\rm{\bullet}$ \emph{NGC\,5506}:  \citet{Uttley05} and \citet{Kelly11} reported on the PSD of NGC\,5506 using \emph{RXTE} data. The PSD bend frequency estimated in the present analysis is a factor $\approx 3$ higher. This may in part be due to the use of different energy bands between the analyses (e.g. 2--10 keV for \emph{RXTE} and 0.2--10 keV \emph{XMM-Newton}).

$\rm{\bullet}$ \emph{NGC\,5548}:  \citet{Markowitz03} and \citet{Kelly11} report on the PSD of NGC 5548. They found a break/bend frequency at $\log(\nu_{b}) \sim -6.1$, far outside the the frequency bandpass avilable to the \emph{XMM-Newton} data; as expected given these results we find no strong evidence for a bend in the present analysis. The single power-law fit (i.e. Model A) shows a slope consistent with the previous high frequency slope estimates.

$\rm{\bullet}$ \emph{Ark\,564}:   The PSD of this source has been discussed by \citet{Pounds01} and \citet{Markowitz03} using \emph{RXTE} data, \citet{Papadakis02} using \emph{ASCA} data, and \citet{McHardy07} and \citet{Kelly11} using combinations of these datasets. The bend frequency estimate reported in the present paper is a factor $\sim 6$ lower than that reported by \citet{McHardy07}, but consistent with that from \citet{Kelly11}. However, as discussed by \citet{McHardy07} the PSD of this source may be more complicated than the simple bending power law model assumes.

$\rm{\bullet}$ \emph{PKS\,0558-504}:  \citet{Papadakis10}  combined \emph{XMM-Newton} and \emph{RXTE} data to estimate a PSD bend frequency of $\log(\nu_{b}) \approx -5.2$, outside the the frequency bandpass avilable to the \emph{XMM-Newton} data. The apparent detection of a PSD bend, as reported in table 3 is therefore quite unexpected. However, \citet{Papadakis10} found evidence for a bump (perhaps a broad QPO) in the PSD at frequencies consistent with our bend. Thus, our bend detection is plausibly a misidentification caused by the more complicated PSD shape.

$\rm{\bullet}$ \emph{NGC\,7469}:  The 2--10 keV PSD of NGC\,7479 was discussed by \citet{Markowitz10} using \emph{RXTE} and \emph{XMM-Newton} data. The found a bend outside the frequency bandpass available to the \emph{XMM-Newton} data. As expected based on this, we find no evidence for a bend using the \emph{XMM-Newton} data alone. The high frequency slope reported by \citet{Markowitz10} is consistent with our slope using the single power-law model ($\alpha = 1.8 \pm 0.2$). 
% =============================================================================
% =============================================================================

%\newpage
%\onecolumn
\section*{Appendix B: Catalogue of PSDs}
(online material)
\end{appendix}

% =============================================================================
% =============================================================================

\end{document}